\documentclass{ifacconf}

\usepackage{graphicx}      % include this line if your document contains figures
\usepackage{natbib}        % required for bibliography
\begin{document}
\begin{frontmatter}

\title{Power Generation Nowcasting of the Behind-the-Meter Photovoltaic Systems} 

\author[First]{Soheil Pouraltafi-kheljan} 
\author[Second]{Murat Göl}

\address[First]{Department of Electrical and Electronics Engineering, Middle East Technical University, Ankara, Turkey (e-mail: soheil.kheljan@metu.edu.tr).}
\address[Second]{Department of Electrical and Electronics Engineering, Middle East Technical University, Ankara, Turkey(e-mail: mgol@metu.edu.tr)}

\begin{abstract}                % Abstract of not more than 250 words.
Rapid increase in the number of photo voltaic systems (PVSs) across the low voltage distribution systems required real-time monitoring of those systems. However, considering the necessary investment cost, real-time monitoring infrastructure installation does not seem to be achieved in the near future.  Thus, alternative approaches to achieve this necessity is crucial. In this study, a method is proposed to nowcast the power generation of the behind-the-meter PVSs. The method relies on the strong correlation between the harmonic current injection of the PVS inverter and the power generation of the PVS. The Artificial Neural Network method (ANN) is utilized to model the relation between the predictors and the aggregated PVS output, i.e. generated power. Finally, the effectiveness of the approach is assessed with real field data.
\end{abstract}


\end{frontmatter}
%===============================================================================

\section{Introduction}
In recent decades, due to the fast-growing electricity needs across the world, different types of distributed energy resources (DERs) penetrated to the conventional power grids.  Consequently, unexampled changes are observed in power systems. The roof-top photo voltaic systems (PVS), which are installed behind-the-meter, constitute a major challenge for today’s distribution system operators as they are not monitored. PVS power output may fluctuate frequently due to moving clouds. Generation fluctuation of a single PVS may not matter to the system operator. However, in the presence of many PVSs, utilities or system operators may suffer from this variability (\cite{Bayram2017:1,Cohen2015}). Moreover, penetration of the PVSs into the grid contributes to the masked load of a feeder which causes a lack of accurate monitoring over the actual load of the feeder. Achieving observability to overcome mentioned dilemmas requires significant investment into monitoring and metering infrastructure.\par
This paper presents a method to nowcast the aggregated power generation of PVSs connected to the same feeder. The method utilizes the current harmonic measurements taken from the considered feeder, along with the temperature and irradiation measurements taken at the same location. Then this information is given to Artificial Neural Networks (ANN) based model as inputs to nowcast the aggregated power generation of the PVSs connected to the considered feeder. Note that, the proposed method is not a forecasting method, rather it aims to determine the power generated by the PVSs at the considered region corresponding to the instant that measurements are taken. \par
The efficiency of the photo voltaic (PV) panel is affected by temperature which results in variations in the power output level of the panel. Moreover, PVS output is highly related to solar irradiation. However, the prediction of total power generation based on the solar irradiation measurements may be misleading as cloudiness over the considered region may not have the same density over the region. Therefore, one needs additional measurements to predict the total solar generation in the considered region. PVS panels are interfaced to the grid via inverters, which are harmonic current sources. Despite, PVS harmonic current emission undeniably is related to the inverter’s control strategy and technology, harmonic current emission patterns of the various PV inverters are almost similar. In our practical observations, the correlation between the PVS generation and the magnitude of the harmonic currents is recorded. Assessment of the observations revealed that harmonic measurements can be used to provide the measurement redundancy along with irradiation and temperature measurements.
The major problem in the utilization of the current harmonics is the presence of other harmonic sources in the grid except for PVSs. The proposed method considers three-phase PVSs and uses three-phase measurements to differentiate harmonics due to three-phase PVSs from the harmonics due to the single-phase harmonic sources of the distribution system.\par
The paper is organized as follows, in the literature review is given in Section II. Section III presents the proposed method, and the results are shown in Section IV. Finally, conclusions are given in Section V.
\section{Literature review}
Presented works in literature on PVS power generation prediction and estimation can be categorized into two sets. The first set includes model-based approaches, which highly depends on the estimation of PV parameters and meteorological parameters. The second set consists of data-driven methods based on various measurement data from different sources. \par
In model-based methods, Direct Normal Irradiance (DNI) and Direct Horizontal Irradiance (DHI) are extracted from measured Global Horizontal Irradiance (GHI). In those methods generated power is estimated using considered PV model, and DHI and GHI. \cite{Hong2014} analysis to detect the impact factors of PVSs. \cite{Gotseff2014}are developed a GIS-based optimization model to estimate power generation of the PVSs. The authors performed a sensitivity   extracted global, direct and diffuse radiation components from global irradiance by implementing an irradiance pre-processing method. This information is taken as input to the SAM software package to estimate the injected solar power. \cite{Chen2017} used the clear sky generation model associated with the physical characteristics of PV to estimate PVS generation. Meteorological data, accurate geographic information, and precise PV characterization of arrays are essential for the model-based scheme. \cite{Jamaly2013} utilize satellite data to derive irradiation data which is applied to the PV performance model. In case of behind-the-meter PV failure, overestimation error could occur in the model-based approaches (\cite{Zhao2015}).\par
In the data-driven approaches, solar generation is disaggregated from net load demand by employing various data collected form the sources such as micro-phasor measurement units ($\mu$PMUs), supervisory control and data acquisition infrastructure (SCADA) and smart meters. \cite{Shaker2016} proposed an up-scaling method based on the recognition of a subset of the solar sites whose data could reveal useful information for estimation of the total power generation from a much larger set of sites. Authors have implemented a hybrid data dimension reduction approach based on k-means and PCA methods to define a subset of sites. Four mapping functions are explored and compared by the authors. Power generation and location information (longitude and latitude) of each solar site and some other measurements such as temperature, cloud coverage, etc. for three to four months' worth of data with a resolution of 15-min have been used for model training. In their further work \cite{Shaker2016:2} have proposed an approach based on their previous work \cite{Shaker2016} utilizing unsupervised learning. To deal with uncertainties of PVS power generation, they have chosen fuzzy arithmetic. This model relies on publicly available data. Some other researchers have proposed data-driven methods to disaggregate the solar power generation from the net load profile. \cite{Sossan2018} presented unsupervised disaggregation of photo voltaic production from measurements of the aggregated power flow at the point of the common coupling and local GHI measurements. Authors consider PVS generation as a function of the GHI which enables them to identify PVS generation patterns in aggregated power flow measurements. They have shown and explained that there are similarities between the time series of the aggregated estimated PVS generation both in time and frequency domains. It is mentioned that the dynamics of the power flow at the point of the common coupling are influenced by PVS generation in a certain frequency range. \cite{Kara2018} demonstrated a set of methods to disaggregate PVS active power generation from the active load using feeder-level $\mu$PMU measurement data and solar irradiance measurements. The authors proposed linear regression to predict actual load using measured reactive power, and the metered nearby PVS generation. A modified version of Contextually Supervised Source Separation (CSSS) methodology is applied to disaggregate the active power measured at the feeder head as the PVS active power generation and load active power demand. Some limitations for this approach are mentioned by the authors such as the necessity of inverter participation in reactive power control. In addition, the performance of the presented approach is only examined for only a centralized large PVS.\par
In addition to the mentioned approaches, hybrid procedures were also adopted. \cite{Bright2018} developed an approach utilizing satellite-derived GHI estimations associated with a varying number of observable reference PVSs to nowcast the PV power generation. A correction factor for each reference site is derived by spatial interpolation of the difference among the measured PVS output and the predicted power by satellite-derived GHI estimations. Then, the correction factor itself is spatially interpolated which applied to the satellite-derived PVS generation estimator. 
\section{The proposed method}
Most of the PVS penetration in low voltage distribution systems is behind-the-meter, so injected power by these solar sites are not metered. This lack of visibility cause problems in the forecast of the actual load, which is extremely important in system operation.\par
The uncertain fluctuation of PVS generation may cause voltage regulation problems in distribution systems. Having an accurate enough knowledge about the near-term generation of the PVS contributes to the fewer operations of the tap changing transformers to maintain the voltage within an acceptable interval \cite{Mehmood2018}. Uncertainty related to the fast-moving and fast-changing clouds imposes not only sharp ramp-up and down to the PVS generation, but also, persuasive difficulties to PVS generation prediction and forecasting work.\par
This paper presents a novel method to nowcast the aggregated solar generation of PVSs connected to a feeder, which utilizes the current harmonic measurements taken from the considered feeder head, along with the temperature and irradiation measurements taken at the same location. The proposed method employs an ANN-based model, which is trained by the synthetic scenarios generated by the Monte Carlo simulations.\par 
The features of the ANN-based method are determined by examining the features affecting PVS generation. Although there are many factors affecting PVS generation, only the most significant ones namely temperature and irradiance (\cite{Dincer2010}) are considered in this work. These values are only measured at the feeder head, to reduce the infrastructure cost.\par
The inverters connecting the PV panels to the grid are the source of harmonic current in distribution networks. The harmonic current emission patterns of the different inverter with different strategies are studied widely (\cite{Chicco2009,Macedo2009,Schlabbach2008,Fekete2012,Infield2004}). In the representative  measurements harmonic currents are considered up to 23rd harmonics, Since the harmonic components in the output current caused by switch harmonic are high frequency harmonics and independent of the power out level \cite{Du2013}. Results revealed the correlation between the generated power and harmonic currents of inverter output current. Correlation coefficients between the generated power and output current’s harmonic components’ amplitude of the inverters are given in Table~\ref{tb1} Where coefficients are computed according to the equation ~(\ref{e1}):
\begin{equation} \label{e1}
corr(X,Y)= r_xy ={{\sum_{i=1}^{n} (x_i -\bar{x})(y_i -\bar{y})}\over{\sqrt{\sum_{i=1}^{n} (x_i -\bar{x})^2\sum_{i=1}^{n} (y_i -\bar{y})^2}}}
\end{equation}

\begin{figure} [hbt]
\begin{center}
\includegraphics[height=5.5cm]{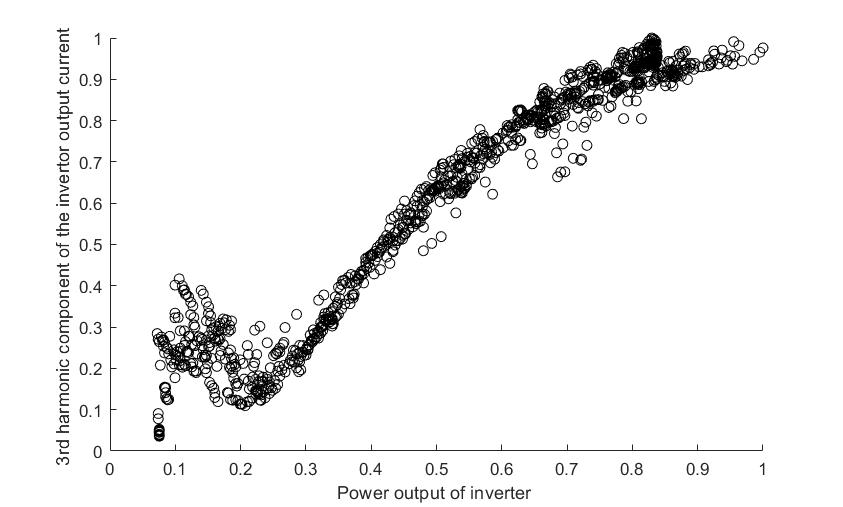}    % The printed column  
\caption{Relation between the 3rd harmonic component of the current and power output of the inverter.}  
\label{fig1}                                 % Size the figures 
\end{center}                                 % accordingly.
\end{figure}

Where $\bar{x}$ and $\bar{y}$ are the averages of the observations of variable x and y respectively. According to Table~\ref{tb1}, the 3rd harmonic component of the inverter output current is highly correlated with the generated power output of the inverter. In other words, the 3rd harmonic component of the inverter output current decreases as the generated power output decreases and vice versa. To make it explicit, the normalized 3rd harmonic component of the inverter output current versus normalized generated power output of the inverter in for 2 days during the day time is depicted in Fig.~\ref{fig1}.

\begin{table}[]
\begin{center}
\caption{Correlation coefficients between harmonic current amplitude and PVS power output
}\label{tb1}
\begin{tabular}{|c|c|c|c|}
\hline
\textit{\textbf{\begin{tabular}[c]{@{}c@{}}Harmonic \\ order\end{tabular}}} & \textit{\begin{tabular}[c]{@{}c@{}}Correlation\\ coefficient\end{tabular}} & \textit{\textbf{\begin{tabular}[c]{@{}c@{}}Harmonic \\ order\end{tabular}}} & \textit{\begin{tabular}[c]{@{}c@{}}Correlation\\ coefficient\end{tabular}} \\ \hline
\textbf{1st}                                                                & 1                                                                          & \textbf{13th}                                                               & 0.79                                                                       \\ \hline
\textbf{2nd}                                                                & -0.32                                                                      & \textbf{14th}                                                               & -0.21                                                                      \\ \hline
\textbf{3rd}                                                                & 0.98                                                                       & \textbf{15th}                                                               & -0.43                                                                      \\ \hline
\textbf{4th}                                                                & -0.72                                                                      & \textbf{16th}                                                               & -0.48                                                                      \\ \hline
\textbf{5th}                                                                & 0.36                                                                       & \textbf{17th}                                                               & -0.49                                                                      \\ \hline
\textbf{6th}                                                                & -0.59                                                                      & \textbf{18th}                                                               & -0.06                                                                      \\ \hline
\textbf{7th}                                                                & 0.1                                                                        & \textbf{19th}                                                               & 0.21                                                                       \\ \hline
\textbf{8th}                                                                & -0.6                                                                       & \textbf{20th}                                                               & -0.41                                                                      \\ \hline
\textbf{9th}                                                                & -0.10                                                                      & \textbf{21st}                                                               & -0.62                                                                      \\ \hline
\textbf{10th}                                                               & -0.28                                                                      & \textbf{22nd}                                                               & -0.81                                                                      \\ \hline
\textbf{11th}                                                               & 0.85                                                                       & \textbf{23rd}                                                               & 0.20                                                                       \\ \hline
\textbf{12th}                                                               & -0.25                                                                      &                                                                             &                                                                            \\ \hline
\end{tabular}
\end{center}
\end{table}

Although, training of the ANN is performed considering the 3rd harmonic component of the inverter output current, determination of pure 3rd harmonic components injected by the PVS to the grid seems to be not possible. To deal with this challenge, 3rd harmonic of the zero sequence of the measured current at the head of the feeder is given as the input to the model to estimate the aggregated generated power of the behind-the-meter PVSs connected to the feeder.\par
In this manner, harmonic currents injected to the grid by single phase loads are filtered out in zero sequences current component, since injected single phase harmonic are unbalanced. In our observation and simulations, a high correlation between the 3rd harmonic component of the inverter output current and the 3rd harmonic component of the zero sequence of the current at the feeder is recorded.

In this manner, harmonic currents injected to the grid by single phase loads are filtered out in zero sequences current component, since injected single phase harmonic are unbalanced. In our observation and simulations, a high correlation between the 3rd harmonic component of the inverter output current and the 3rd harmonic component of the zero sequence of the current at the feeder is recorded.\par
Furthermore, the optimal installation angel of PV  also affects the PV generation which depends on the location of the PV \cite{Zhao2010}. However, the proposed method aims to determine the power generation of PVSs on the same feeder, their locations are close, and accordingly, their optimal installation angles are almost the same. However, as it does not change with time, it is not included as a feature.
\subsection{ANN}
ANN is a powerful nonlinear data driven computational tool, which is modeled easily and less time consuming than other mathematical methods for the complex real-life problems. A class of feed-forward ANN, which consists of multiple layers of neurons is a multi-layer perceptron (MLP). Generally, the MLP’s structure is composed of at least three main layers including input, hidden and output layers. The intermediate neurons connecting the input and output layers are known as hidden, as those neurons do not interface with the external environment. Despite there is no any specific logic to define the the architecture of MLPs, as an often cited theoretical finding, \cite{bengio2016deep} have shown that any function can be approximated by a MLP with single hidden layer .In our approach, MLP is utilized to map the predictors to the dependent variable, which is generated power of the behind-the-meter PVS. Since we are dealing with the regression problem, the activation function of neurons in all layers is defined to be a rectified linear unit (ReLU). Moreover, Adam optimizing algorithm is implemented to optimize the weights of the MLP to achieve the optimum. Number of the nodes in the single hidden layer (150), and the batch size (33000) are defined through a randomized  grid search.

\begin{figure} [hbt]
\begin{center}
\includegraphics[height=15cm]{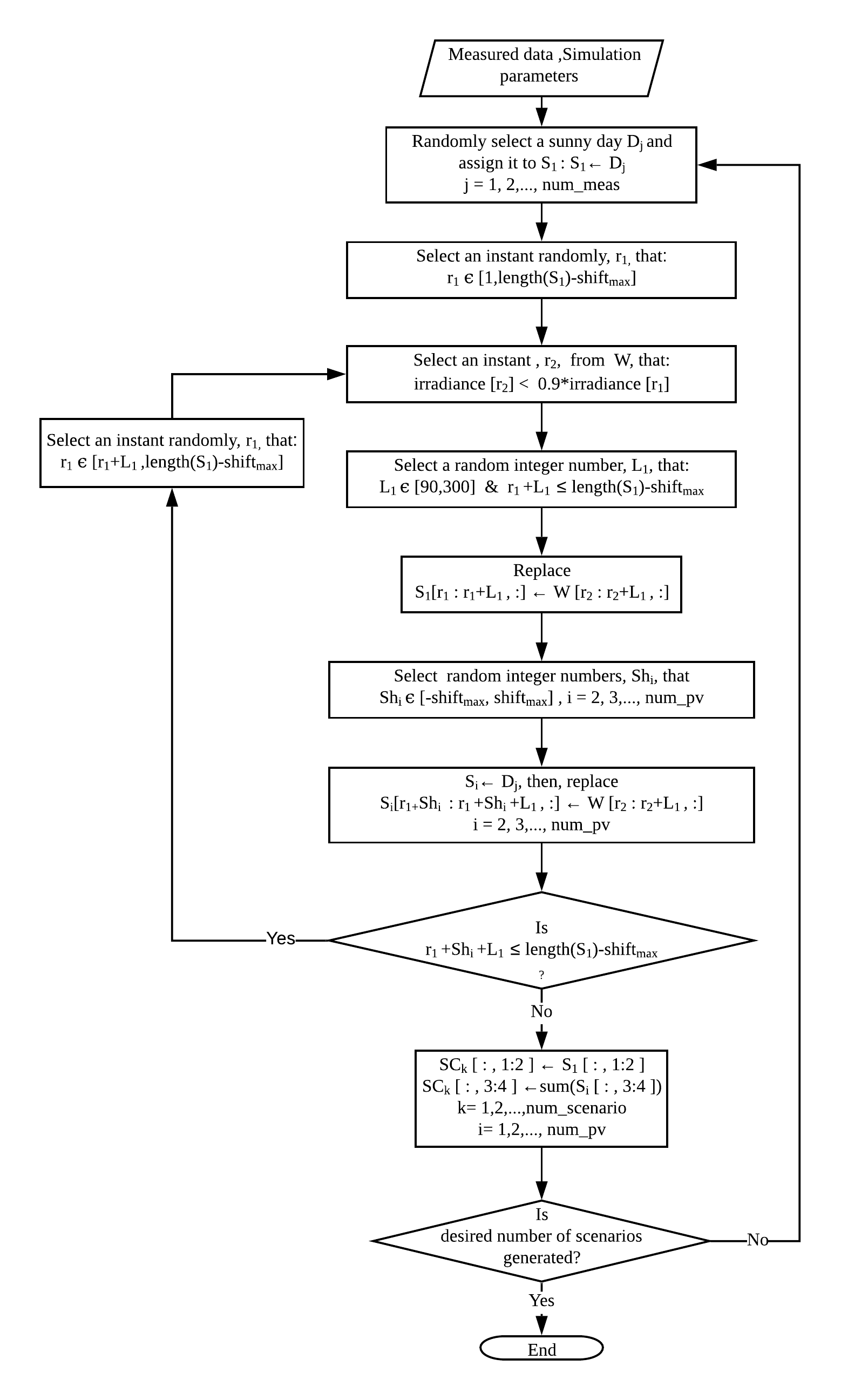}    % The printed column  
\caption{Flow chart of the training scenario generation.}  % width is 8.4 cm.
\label{fig2}                                 % Size the figures 
\end{center}                                 % accordingly.
\end{figure}

\subsection{Training of the Model with Monte Carlo Simulations}
The clouds' motion may cause the intermittency of solar irradiance, which is the main problem of the PVS generation nowcasting. Tracking and predicting such a movement along with forecasting irradiance make a dilemma in PV generation estimation. However, utilizing machine learning techniques could pave to overcome this challenge. On the other hand, adequate training data to make the algorithm learn is necessary. Because of the lack of proper monitoring infrastructure, one cannot collect data synchronously from all PVSs connected to the same feeder. Therefore, Monte Carlo simulations are performed to synthesize various scenarios using the real data corresponding to one PVS. The considered features, namely the temperature of the medium, irradiance, the amplitude of the 3rd harmonic component of the output current at the inverter terminal, and the generated power of real PVS are measured for fourteen days with the resolution of 1 second. Within this period, rainy, cloudy, partly cloudy, and sunny days are experienced. These measured real data are organized as the data set, $W$, including four columns where each column is assigned to a measured feature. Temperature, irradiance, 3rd harmonic component of the current, generated power corresponds to the column respectively. Moreover, the corresponding measurement instances of day i are assigned to $D_j$ where j=1,2,…,num\_meas and, num\_meas is the number of the days within the measurement period.\par
In the Monte Carlo simulations, it is assumed that the irradiance decrease in 1 second should be less than 90\% of the irradiance of the considered instance. Moreover, it is assumed that the duration of cloud coverage should be more than 1.5 minutes. Finally, it is assumed that maximum latency for cloud coverage over the different PVSs connected to the feeder is 300 seconds (5 minutes) which is applied to simulation by the $shift_{max}$ parameter.\par 
In order to generate a synthetic cloudy day scenario, first, a data-set related to a sunny day is randomly selected and then it’s assigned to $S_1$ in which other measured instances will be replaced.
On the next step, some instances of the $S_1$ will be randomly replaced by other randomly selected instances from the $W$ regarding the mentioned assumptions above. In the case of multiple PVSs, first, the same sunny day $D_j$ will be copied to the $S_i$ where i=2,3,…,num\_pv, and num\_pv is the number of PVSs in the proposed system.

\begin{figure} [hbt]
\begin{center}
\includegraphics[height=16 cm]{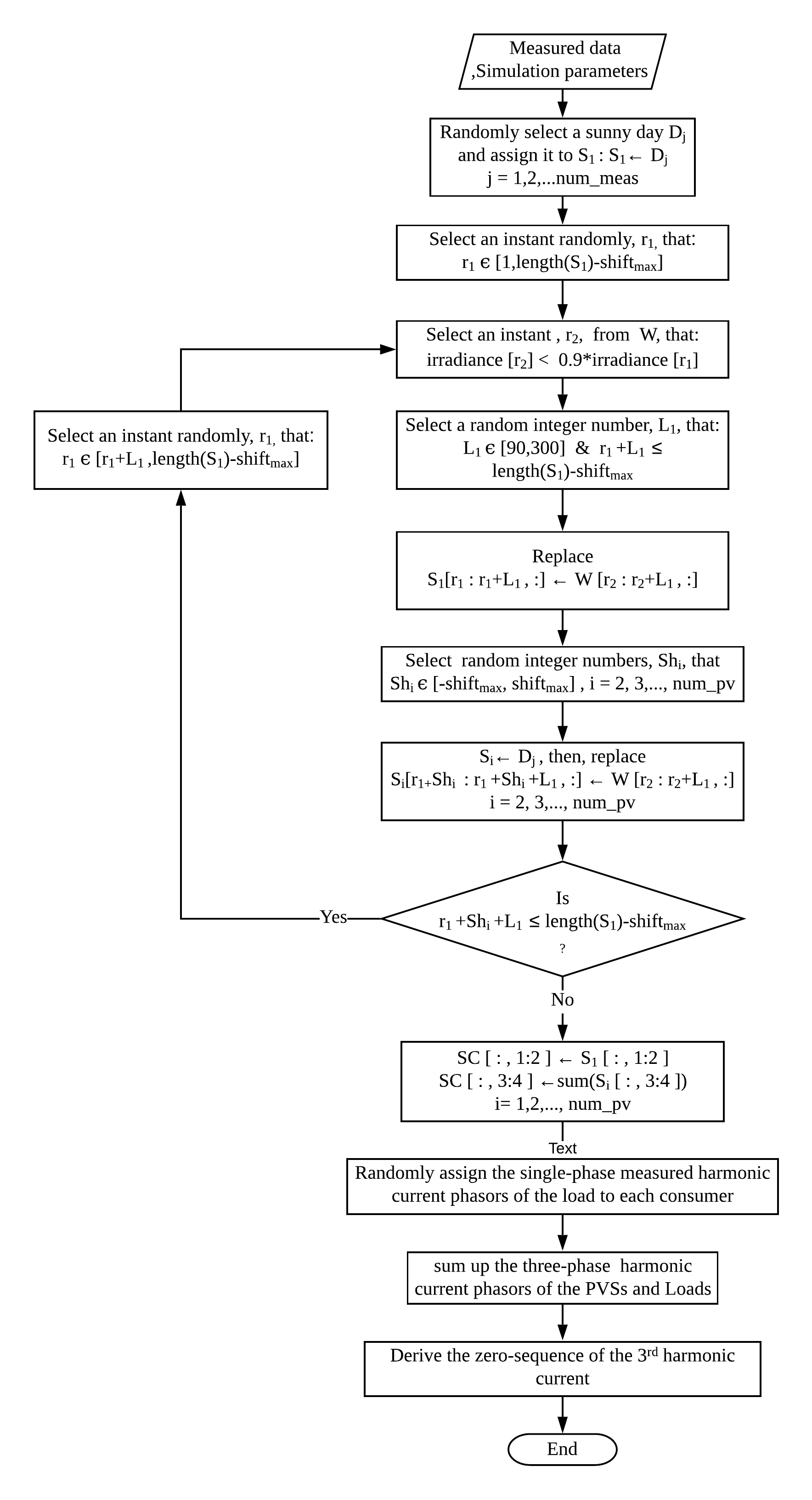}    % The printed column  
\caption{Flow chart of the testing scenario generation.}  % width is 8.4 cm.
\label{fig3}                                 % Size the figures 
\end{center}                                 % accordingly.
\end{figure}

But then, the replacing instances from the W will be shifted backward of forward. Note that all the added data points are real measured values of the temperature, irradiance, generated power, 3rd harmonic component of the inverter output current, but these values are just displaced. Finally, despite, the corresponding generated power and 3rd harmonic current of  $S_i$ , i=1,2,3,…,num\_pv are summed up to achieve aggregated generated power at the head of the header for a synthesized scenario. The corresponding temperature and irradiance of the $S_1$ is taken as the temperature and irradiance of this scenario. The detailed illustration of the discussed algorithm is presented in Fig.~\ref{fig2}. 

This process is repeated for several times to generate different cloudiness scenarios. In this work, 1500 different scenarios are synthesized to train our ANN-based model.\par
Another algorithm is developed to generate synthetic test scenarios to assess the performance of the presented method shown in Fig.~\ref{fig3}. In this manner, the presence of the other harmonic resources in the grid are considered. Like the previous algorithm, power generation and harmonic emission patterns of the PVSs for cloudy days are synthesized, and then emitted 3rd harmonic current of single-phase loads of the consumers is added in a way that, different harmonic profiles are assigned to each phase of the consumer load. Finally, the zero-sequence operation is applied to the aggregated harmonic current. Note that only the injected harmonics of the loads are considered since the fundamental current of these loads has no effect on the process.

\section{Results}
The performance of the proposed method is analyzed in two different cases. In both cases time resolution is 1 second. Utilized data in the first case is collected from the PVS located on top of the Ayasli Research Center at the Middle East Technical University. For the second case, a synthetic scenario is generated with multiple solar sites.

\subsection{Case a: Single PVS connected to the feeder}
\begin{figure} [hbt]
\begin{center}
\includegraphics[height=5.5cm]{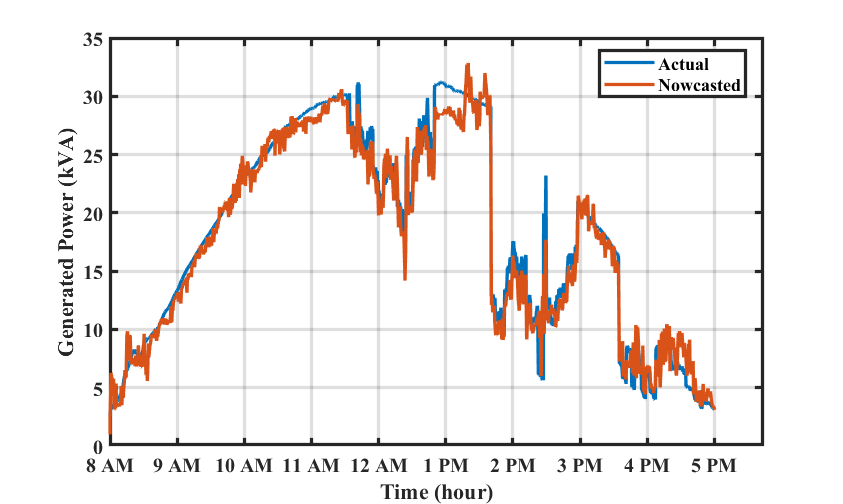}    % The printed column  
\caption{Comparison of the actual and nowcasted values for Case a.}  % width is 8.4 cm.
\label{fig4}                                 % Size the figures 
\end{center}                                 % accordingly.
\end{figure}

In this case, it is considered that only a single PVS is connected to the feeder alongside the loads. It should be noted that the correlation coefficient between the 3rd harmonic component of the inverter output current and the 3rd harmonic component of the zero sequence of the current at the feeder head is 0.96 which shows the that unbalanced 3rd harmonic components of current injected by single phase costumer's loads affect the zero-sequence component of current at the head of the feeder insignificantly. Resultant estimations for a cloudy is presented in Fig.~\ref{fig4}, where RMSE value and mean absolute error of the estimated values are 1.57 kVA (5.00\%) and 1.22 kVA (3.91\%) respectively.

\subsection{Case b: multiple PVSs connected to the feeder}
\begin{figure} [hbt]
\begin{center}
\includegraphics[height=4.3cm]{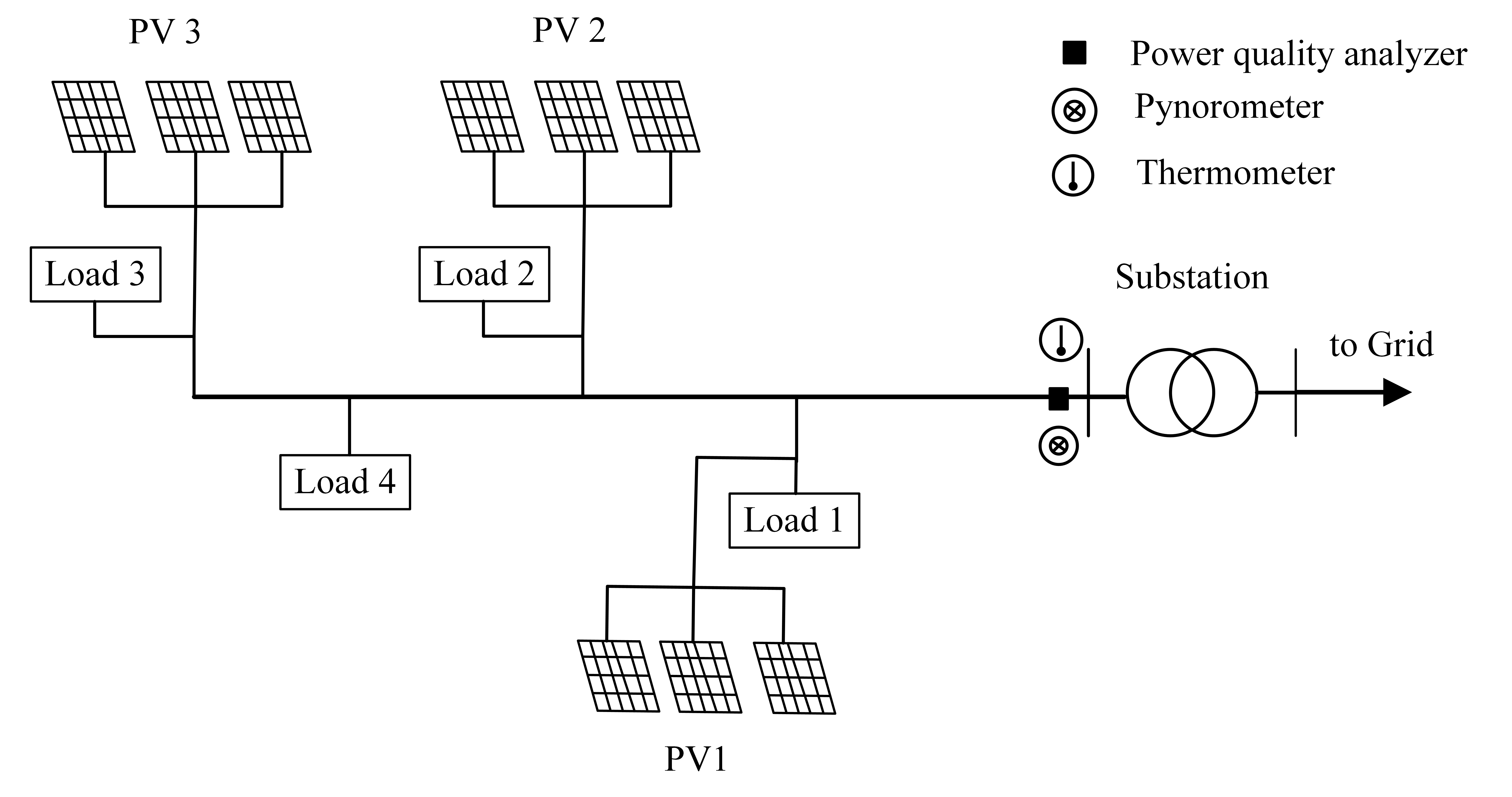}    % The printed column  
\caption{Structure of system in Case b: multiple PVSs connected to the feeder.}  % width is 8.4 cm.
\label{fig5}                                 % Size the figures 
\end{center}                                 % accordingly.
\end{figure}

In the second case, a system with of multiple PVSs is considered. Three PVSs and four costumer loads are connected to the feeder, which is shown in Fig.~\ref{fig5}. Because of the different locations of the PVSs, latency in cloud motion and partial cloudiness is experienced in this case. The mean absolute error and RMSE values are 2.02kw (2.08\%) and 2.77kw (2.85\%) respectively.

\begin{figure} [hbt]
\begin{center}
\includegraphics[height=5.5cm]{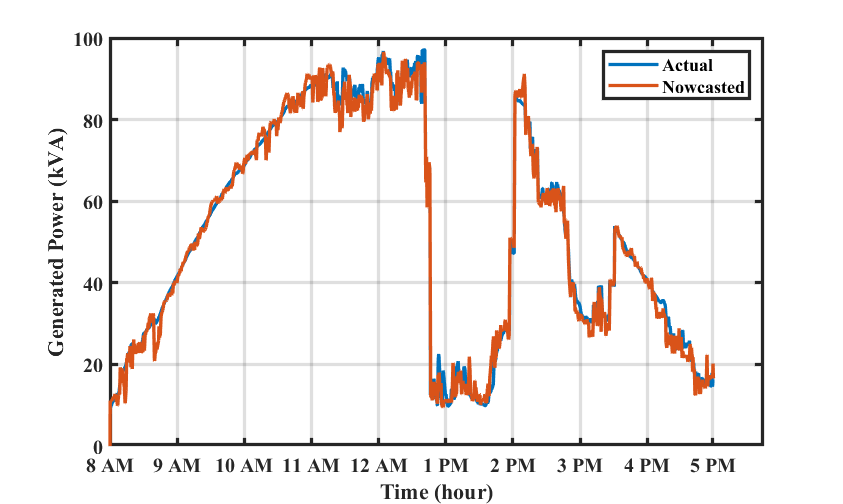}    % The printed column  
\caption{Comparison of the actual and nowcasted values for Case b.} 
\label{fig6}                                 % Size the figures 
\end{center}                                 % accordingly.
\end{figure}

\section{Conclusion}
Rapid integration of PVSs in distribution systems imposed challenges for distribution system operators. It is essential to have enough monitoring over the generation of PVSs to overcome these challenges, which seems to be impractical in short term due to the cost of the monitoring infrastructure installation. Therefore, an ANN based method is developed for nowcasting of the behind-the-meter PVS generation connected to a feeder, by utilizing the harmonic currents at the head of the feeder. The main motivation of this approach is the observed high correlation between PVS power generation and harmonic components of the PVS inverter current. To deal with the uncertain and fast-moving cloud coverage over the area, an ANN model is trained by the synthetic scenarios generated by Monte Carlo simulations. To filter the harmonic injection to the feeder from the loads connected to the feeder, consistency among the three-phase measurements is utilized. 
\bibliography{ifacconf}             % bib file to produce the 
\end{document}